\documentclass[aps,prl,letterpaper,11pt,twoside,tightenlines,nofootinbib,showpacs,preprint,twocolumn]{revtex4}
\usepackage{graphicx}
\usepackage[sort&compress]{natbib}
\begin{document}
\newcommand{\eg}{{\it e.g.}}
\newcommand{\etal}{{\it et. al.}}
\newcommand{\ie}{{\it i.e.}}
\newcommand{\be}{\begin{equation}}
\newcommand{\dd}{\displaystyle}
\newcommand{\ee}{\end{equation}}
\newcommand{\bea}{\begin{eqnarray}}
\newcommand{\eea}{\end{eqnarray}}
\newcommand{\bef}{\begin{figure}}
\newcommand{\eef}{\end{figure}}
\newcommand{\bce}{\begin{center}}
\newcommand{\ece}{\end{center}}
\def\lsim{\mathrel{\rlap{\lower4pt\hbox{\hskip1pt$\sim$}}
    \raise1pt\hbox{$<$}}}         
\def\gsim{\mathrel{\rlap{\lower4pt\hbox{\hskip1pt$\sim$}}
    \raise1pt\hbox{$>$}}}         

\title{ Effective degrees of freedom of the Quark-Gluon Plasma}
\author{P.~Castorina}
\affiliation{Dipartimento di Fisica, Universit\`a di Catania,
 and INFN Sezione di Catania, Via Santa Sofia 64
I-95100 Catania, Italia}

\author{M.~Mannarelli}
\affiliation{Center for Theoretical Physics, Laboratory for Nuclear
Science and Department of Physics
\\Massachusetts Institute of Technology, Cambridge, MA 02139.}
\date{\today}
\begin{abstract}
The effective degrees of freedom of the Quark-Gluon Plasma  are
studied in the  temperature range $\sim 1-2$ $ T_c$. Employing
lattice results for the pressure and the energy density, we
constrain the quasiparticle chiral invariant mass to be of order
$200$ MeV and the effective number of bosonic resonant states to be
at most of order $\sim 10$. The chiral mass and the effective number
of bosonic degrees of freedom decrease with increasing temperature
and at $T \sim 2$ $T_c$ only quark and gluon quasiparticles survive.
Some remarks regarding the role of the gluon condensation and the
baryon number-strangeness correlation are also presented.
\end{abstract}
\preprint{MIT-CTP-3698} \maketitle

Quantum Chromodynamics (QCD) predicts that at extremely high temperatures
 matter consists of a gas of weakly interacting quarks and gluons, the
Quark-Gluon Plasma (QGP).  However at moderate temperatures $T= 1 -
2$ $T_c$,  ($T_c = 170$ MeV  deconfinement temperature) it is less
clear what  the dynamical phase is.

The experimental data obtained at the
Relativistic-Heavy-Ion-Collider (RHIC), with the measurement of the
$p_t$ spectra and  the related indications on the radial and
elliptic flow (see \cite{Kolb:2003dz} for a review), clearly suggest
that at moderate temperatures the produced system is in a strongly
interacting phase (sQGP) and  there are remnant of the confining
interaction  up to temperatures $\simeq 2$ $T_c$,  in agreement with
lattice (lQCD) results. Actually lattice  calculations of the
pressure and energy density of the system do not reach the
Stefan-Boltzmann values for a weakly interacting quark-gluon plasma
even at very large temperatures $\simeq 5$
$T_c$~\cite{Karsch:2003jg,Karsch:2000ps}.

This surprising picture calls for understanding the relevant degrees
of freedoms to describe such a phase and, in this respect, several
models have been proposed, where deconfinement and chiral symmetry
restoration occur at a lower temperature than the $q \bar q$
dissociation temperature and ``resonance" states may play an
important dynamical role
\cite{Hatsuda:1985eb,Shuryak:2004tx,Mannarelli:2005pz,Mannarelli:2005pa,Castorina:2005tm,vanHees:2005wb}.

However, recent  lattice and phenomenological analyses
\cite{Koch:2005vg} have shown that the emerging degrees of freedom
are quark and gluon quasiparticles and this result has to be
compatible with the survival of light  $\bar q q$
states~\cite{Karsch:2003jg,Asakawa:2003nw} as obtained by the
analysis of mesonic spectral functions above $T_c$.

In this letter we address the previous puzzling aspects by
performing a phenomenological analysis of  the pressure and of the
energy density of the system  taking into account the presence of
quark and gluon quasiparticles as well as of $\bar q q$ states in
the temperature range $ 1 - 2 $ $T_c$.

Let us consider a system of quark, antiquark, gluons and  correlated
particle states.

The gluonic sector contains quasiparticle contributions  as well as
non perturbative condensation effects. The ratio, $c_e(T)$, between
the chromo-electric condensate evaluated at finite temperature and
at $T=0$ strongly decreases for temperature $T > 1.2$ $T_c$, whereas
the same ratio for the magnetic condensate is $\simeq 1$ up to
larger temperatures \cite{D'Elia:2002ck}. Since at $T=0$ the
chromo-electric and chromo-magnetic parts are equal, one can write,
in the considered temperature range, the gluon condensate as
\begin{equation}
<\frac{\alpha_s}{\pi} G^a_{\mu \nu} G_a^{\mu \nu}>_T = \frac{1}{2}
<\frac{\alpha_s}{\pi} G^a_{\mu \nu} G_a^{\mu \nu}>_0 [ 1 + c_e(T)]
\label{gluocon}\end{equation} where the ratio  $c_e(T)$ can be
approximated by the unquenched data of Ref. \cite{D'Elia:2002ck} and
we take $ <\alpha_s/\pi G^2>_0 \simeq 0.01$ GeV$^4$, consistently
with  QCD sum rules \cite{Novikov:1983jt}.

The gluon condensate, i.e. a macroscopically
populated state with zero momentum, does not essentially contribute
to the pressure but is crucial for the evaluation of the energy
density \cite{Zwanziger:2004np}. On the other hand, the gluonic sector contains also gluon quasiparticles
which contribute to the pressure and to the energy density and that we shall treat in a phenomenological way
(see below).

Concerning the fermionic sector,  we assume that the number of
quark/antiquark degrees of freedom is fixed, $D_q=D_{\bar q} = 18$.
Recently some dispersion relations of the general form
\begin{equation}
\omega_{\bar q}(k)\,=\,\omega_q(k)\,=\, \sqrt{k^2 + m^2} + \Sigma_R
\label{dispersion}\, ,\end{equation} have been proposed in
\cite{Mannarelli:2005pz,Kitazawa:2005mp} where $m$ and the self
energy $\Sigma_R$ have been  evaluated  taking into account the
interaction of the quasiparticles with the medium.  For the relevant
momenta, of the order of the thermal momentum, we consider that
$m/k$ is small \cite{Mannarelli:2005pz} and treat the chiral
invariant term $\Sigma_R$ as a constant parameter $M$
 by using the dispersion relation
 \be \omega_{\bar
q}(k)\,=\,\omega_q(k)\,=\, k + M \label{dispersion2}\, .\ee
Therefore, at this level of approximation,  we   neglect the
dynamical phenomena related to frequencies with $\omega/T \ll 1 $,
such as viscosity.

The  structure  of the  in-medium correlated states as a function of
temperature is not easily evaluated. These states may describe $\bar
q q$ states as well as more exotic states \cite{Shuryak:2004tx}.
Close to $T_c$, it should be reasonable to consider that the number
of correlated state degrees of freedom $D_{b}$ is of order $10$,
which corresponds to the pseudoscalar nonet. However in our analysis
we will treat $D_{b}$ as a parameter indicating that an effective
number of bosonic states is present (see below). In the following we
will neglect, as a first approximation, the effect on the
thermodynamics quantities of the width of   the bosonic states.
Therefore we employ the dispersion law $ \omega_b = \sqrt{k^2 + 4
M^2}\,$.

Finally, in the present approach, the  interaction of the gluonic sector
with fermions and correlated states is  described, in a mean
field-like treatment, by  the effective mass $M$ and $D_b$ which we
will be evaluated employing unquenched lattice data of pressure and
energy density.

Therefore our expressions of pressure and energy density are given
by the sum of the contributions of quarks, antiquarks, bosonic pairs
and gluons: \be P(T) = \!\!\!\sum_{i=q,\bar q, b}
 \!\!\!D_i\int
\!\!\!\frac{d^3 k}{(2 \pi)^3} \frac{k^2 f_i(k)}{3 \,\omega_i(k)} +
P_g(T) \label{Pressure}\, ,\ee

\be \epsilon(T) = \!\!\!\sum_{i=q,\bar q, b} \!\!\!D_i\int
\!\!\!\frac{d^3 k}{(2 \pi)^3} \omega_i(k) f_i(k) + \epsilon_g(T)+
\epsilon_{ con}(T), \label{energy}\ee where $f_q(k) = f_{\bar q}(k)
$ is the Fermi distribution law, $f_b(k)$ is the Bose distribution
law, $P_g(T)$ and  $\epsilon_g(T)$ are respectively the
contributions to the pressure and energy density due to gluon
quasiparticles and $\epsilon_{ con}(T)$ is  the gluonic condensate
given in Eq.(\ref{gluocon}).

In order to evaluate $M$ and $D_b$ we have to  perform a
simultaneous fit of the data of
Refs.~\cite{Karsch:2003jg,Karsch:2000ps} as a function of the
temperature employing Eqs. (\ref{Pressure}), (\ref{energy}) and
(\ref{gluocon}).

As suggested by quenched data, the contribution, $\epsilon_g$, to
the energy density is small with respect to the gluon condensate and
also the effect of $P_g$  on the pressure of the whole system (i.e.
including fermions and correlated states) is expected to be small
\cite{Karsch:2003jg}.

Therefore, as a first step, let us assume that $P_g=\epsilon_g=0$.
As we shall see, this approximation gives an upper limit on the
effective degrees of freedom above $T_c$. Moreover,
 we will numerically study how
different values of $P_g$ and $\epsilon_g$  may affect our final results.

\begin{figure}[!th]
\includegraphics[width=2.in,angle=-90]{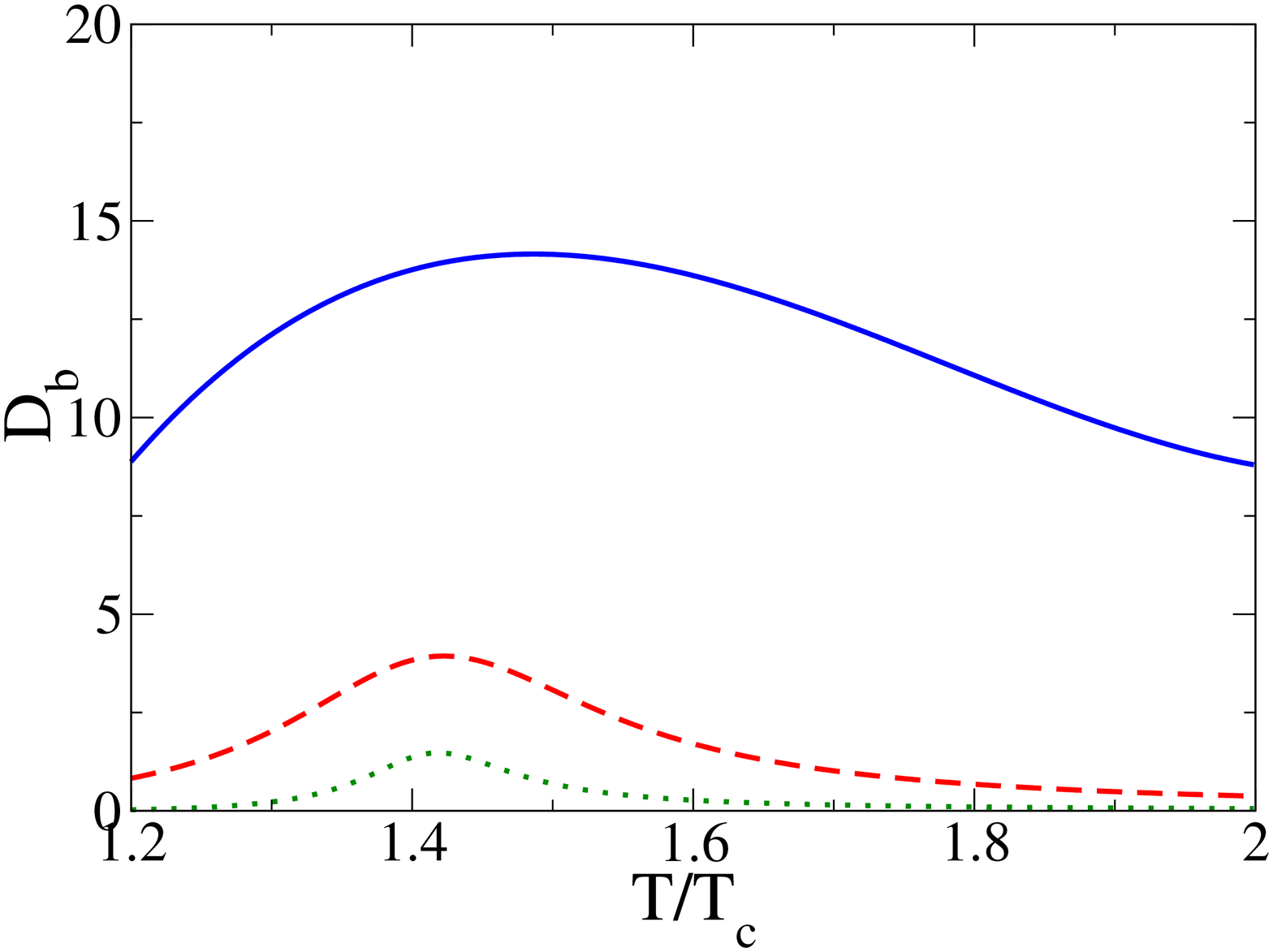}
\includegraphics[width=2.in,angle=-90]{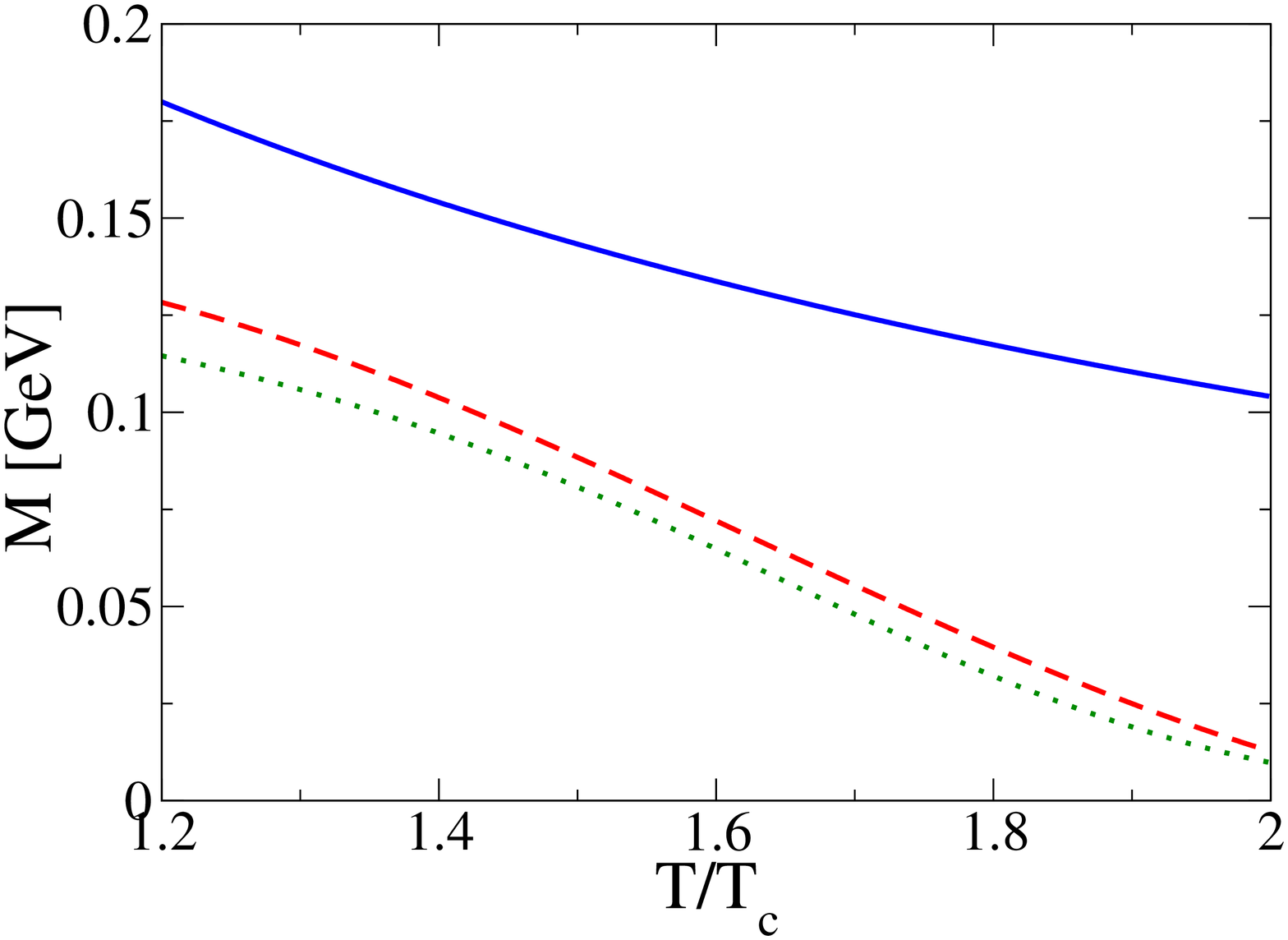}
\caption{Effective number of bosonic degrees of freedom (upper
panel) and quasiparticle chiral mass (lower panel) as a function of
the temperature for $T = 1.2 - 2. T_c$. Full (blue) lines correspond
to $P_b=\epsilon_g=0$; dashed (red) lines correspond to pressure and
energy density evaluated with $m_g=1.0$ GeV and dotted (green) lines
to $m_g=0.9$ GeV.} \label{figNbM}
\end{figure}

The result of the combined, $P(T)$-$\epsilon(T)$, analysis, for
$P_g=\epsilon_g=0$,  are shown in the plots of Fig.\ref{figNbM},
full (blue) lines.

From the upper panel of Fig.\ref{figNbM}  one  sees that  also when
the gluon quasiparticles are turned off, i.e.  one is artificially
increasing the number of  correlated pairs, the effective  number of
bosonic degrees of freedom, in the range $T=1.2 - 2$ $T_c$, remains
of order $10$, suggesting that essentially the  states of the
mesonic nonet are present.

The chiral invariant mass of the quark quasiparticles, shown in the
lower panel of Fig.\ref{figNbM}, turns out to be a decreasing
function of the temperature suggesting that the mechanism which
determines the chiral mass becomes less efficient as the temperature
increase. It is interesting to note that  the decreasing of the
chiral mass as a function  of the temperature determined with this
approach is in qualitative agreement with the one determined in
Ref.\cite{Mannarelli:2005pz} with a different method.

Notice that the previously obtained value of   $D_b$ has to be
considered as an upper limit to the effective number of correlated
degrees of freedom. Indeed, if in Eqs.(\ref{Pressure}) and
(\ref{energy}) one switches on gluons, that is if  one includes the
contributions of the gluon quasiparticles to the pressure and to the
energy density, these terms reduce the weight of the fermions and of
the correlated bosonic states. To check numerically this effect the
same unquenched lattice results have been fitted with $P_g$ and
$\epsilon_g$ calculated with different
 gluon quasiparticle masses $m_g$ and for $D_g=24$ gluonic degrees of freedom.
The results are reported in Fig. \ref{figNbM} which shows that
increasing the values of $P_g$ and $\epsilon_g$, i.e. employing
different values of $m_g$, $D_b$ and $M$ drastically
decrease\footnote{The bump in $D_b$ at $T \sim 1.4$ $T_c$ is due to
have considered a thermal independent gluon mass in $P_g$ and
$\epsilon_g$. Indeed, in Ref.\cite{Levai:1997yx}, a minimum in
$m_g(T)$ is found at approximately the same value of $T$. }.

Therefore  our analysis of the unquenched lattice data strongly
constraints the number of correlated states and of the quasiparticle
masses.

Few comments are now in order. Let us notice that the value of the
mass of the quasiparticles that we obtain is smaller than the one
obtained in Ref. \cite{Levai:1997yx} or in Ref.\cite{Peshier:2005pp}
where $M$ is estimated to be  $\sim 3-4\, T$  from fits of lattice
data. This difference  essentially relies on the fact that we have
considered the dispersion law of Eq.(\ref{dispersion2}) with a
chirally invariant mass, whereas in \cite{Levai:1997yx} and
\cite{Peshier:2005pp} the quasiparticle dispersion law has been
parameterized as $\omega_q = \sqrt{k^2 + M^2}$. Moreover in
Ref.\cite{Levai:1997yx}, in order to reproduce the lattice results,
a (small) bag constant has been employed. In our case  the
contribution of the gluonic condensate and of the mesonic resonant
states play a  crucial role   in determining the correct values of
pressure and energy density.

Finally let us comment on the correlation between baryon number and
strangeness ($C_{BS}$) as an  indication of the effective dynamical
degrees of freedom of the system. The analysis of lQCD
 results performed in \cite{Koch:2005vg} indicates that at $1.5$ $ T_c$ the
 BS correlation is very close to 1.

We can estimate, in an admittedly rough way, the BS correlation as
\be C_{BS}\,\sim\, \frac{\frac{1}{3}D_f <n_f>}{\frac{1}{3} D_f <n_f>
+ \frac{4}{9}D_b<n_b> }
 \, ,\ee
where $<n_b>$ ($<n_f>$) is the number density of bosonic (fermionic)
states, the coefficient $D_f/3$ takes into account that in the
chiral symmetric limit  one third of fermions are strange, whereas
the coefficient $4/9D_b$  is an effective way to weight the number
of strange boson in $D_b$ according to the meson nonet.

Employing the data of Fig.\ref{figNbM} at $T=1.5$ $T_c$ it turns out
that the correlation is about $0.95$ for $m_g =1.0$ GeV. Using
smaller values of  $m_g$ the correlation  further increases.
Moreover in every case considered at $T=2.0$ $ T_c$ the correlation
is in any cases $\simeq 1$.

In conclusion, according to the present work, the relevant  degrees
of freedom in QCD, for temperatures above $T_c$ are $q$,$\bar q$,
$g$ quasiparticles and bosonic states. The effective number of
degrees of freedom associated with the bosonic states is at most of
order $10$ suggesting that only light non exotic states are present.

For $T \gtrsim 2$ $T_c$ the contribution of
mesonic bound states to pressure and energy density is vanishing
small and only quasiparticles are relevant.  Gluon condensation and
its persistence above $T_c$ is a fundamental ingredient of the
energy balance.

 The correlation between baryon number and
strangeness can be understood by considering the
  reduction of the effective bosonic
degrees of freedom  due to the gluon quasiparticle mass. Further
investigations are needed to clarify the underlying non perturbative
dynamics  in terms of resonance scattering, chiral phase
fluctuations  and instantons.

\noindent
{\bf Acknowledgement} \\
We thank K.~Rajagopal for useful comments. The work of  PC has been
supported by the ``Bruno Rossi" program. The work of MM has been
supported by the ``Bruno Rossi" fellowship and by U.S. Department of
Energy (D.O.E.) under cooperative research agreement
\#DE-FC02-94ER40818.

\end{document}